%% file: morello_ichep08.tex
\documentclass[slac_one]{revtex4}
\usepackage{graphicx}
\usepackage{fancyhdr}
\include{macro}

\begin{document}

\title{Charmless $b$-hadrons decays at CDF} 

%

\author{M.~J.~Morello (for the CDF collaboration)}
\affiliation{University and I.N.F.N. of Pisa, Building C, 
Polo Fibonacci, Largo B. Pontecorvo, 3 - 56127 Pisa, Italy}

\begin{abstract}
We present CDF results on the branching fractions and time-integrated direct \CP\ 
asymmetries for \Bd, \Bs\ and \Lb\ decay modes into pairs of charmless charged hadrons 
(pions, kaons and protons). The data-set for these measurements amounts to 1~fb$^{-1}$ 
of $\bar{p}p$ collisions at a center of mass energy $1.96~\rm{TeV}$.
We report on the first observation of the \BsKpi, \Lbppi\ and \LbpK\ decay modes and on 
the measurement of their branching fractions and direct \CP\ asymmetries. 
\end{abstract}

\maketitle

\thispagestyle{fancy}



\section{INTRODUCTION}
The interpretation of the CP violation
mechanism is one of the most controversial aspects of the Standard Model.
Many extensions of Standard Model predict that there are new sources of CP violation,
beyond the single Kobayashi-Maskawa phase in the quark-mixing matrix (CKM). Considerations 
related to the observed baryon asymmetry of the Universe imply that such new sources should exist.
The non-leptonic decays of $b$ hadrons into pairs of charmless charged hadrons 
are effective probes of the CKM matrix and sensitive to these potential new physics effects.
The large production cross section of $b$ hadrons of all kinds at the Tevatron
allows extending such measurements to \Bs\ and \Lb\ decays,
which are important to supplement our understanding of \Bd\ meson decays.

The branching fraction of \BsKpi\ decay mode could be used to measure
$\gamma$~\cite{Gronau:2000md} and the measurement of its direct
\CP\ asymmetry could be a powerful model-independent test 
of the source of  direct
\CP\ asymmetry in the $B$ system \cite{Lipkin:2005pb}. This may provide 
useful information to solve the
current discrepancy between the direct CP asymmetries observed in the neutral
\BdKpi\ decay mode and charged \BuKpi\ decay mode~\cite{PDG08}.
The \Bspipi\ and \BdKK\  decay modes proceed through annihilation and exchange topologies, which
are currently poorly known and a source of significant uncertainty in
many theoretical calculations~\cite{B-N,Bspipi}. 
A measurement of both decay modes would allow a determination of
the strength of these diagrams~\cite{Burasetal}.
CP violating asymmetries in \Lbppi\ and \LbpK\ decay modes may reach significant size $ \mathcal{O}(10\%)$ 
in the Standard Model \cite{Mohanta1}. 
Measurements of asymmetries and branching fractions of these modes would rule out (or allow) some extensions of 
the Standard Model \cite{Mohanta2}.  

Throughout this paper, C-conjugate modes
are implied and branching fractions indicate
\CP-averages unless otherwise stated.


\section{CDF\,II}
CDF\,II is a multipurpose magnetic spectrometer surrounded by
calorimeters and muon detectors~\cite{CDF}. 
A silicon micro-strip detector (SVXII) and a cylindrical drift chamber
(COT) situated in a 1.4 T solenoidal magnetic field
reconstruct charged particles in the pseudo-rapidity range
$|\eta| < 1.0$.
The SVXII consists of five concentric layers
of double-sided silicon detectors with radii between 2.5 and 10.6 cm,
each providing a measurement with 15~$\mu$m resolution in the
azimuthal ($\phi$) direction and 70~$\mu$m along the beam ($z$) direction.
The COT has 96 measurement layers, between 40 and 137 cm in radius, 
organized into alternating axial and $\pm 2^{\circ}$ stereo ``super-layers''.
The transverse momentum resolution is $\sigma_{p_{T}}/p_{T} \simeq
0.15\%\, p_{T}$/(GeV/$c$) and the observed mass-widths are about 14 \massmev\
for $J/\psi\to\mu^+\mu^-$ decays, and about 9 \massmev\ for \Dkpi\ decays.
 The specific energy loss by ionization (\dedx) of charged particles in the COT
is measured from the amount of charge collected by each wire.
An average separation power of 1.5 Gaussian-equivalent standard deviation is obtained in separating 
pions and kaons with momentum larger than 2~\pgev.


\section{MEASUREMENTS OF \Bhh\ DECAYS}

The Collider Detector at Fermilab (CDF) experiment analysed an integrated luminosity  
$\int\Lumi dt\simeq 1$~\lumifb\ sample of pairs of oppositely-charged particles
with $p_{T} > 2$~\pgev\ and   $p_{T}(1) + p_{T}(2) > 5.5$~\pgev,
used to form hadron candidates.
The trigger required also a transverse opening angle $20^\circ < \Delta\phi < 135^\circ$ between the two tracks,
  to reject background from particle 
pairs within the same jet and from back-to-back jets.
In addition, both charged particles were required to originate from
a displaced vertex with a large impact parameter $d_0$ (100 $\mu$m $< d_0(1,2) < 1$~mm), 
while the \bhadrons\ candidate was required to be produced in
the primary $\bar{p}p$ interaction ($d_0< 140$~$\mu$m) and to have travelled a transverse distance
$\L_{T}>200$~$\mu$m. A sample of about 14,500 \Bhh\ decay modes  
(where  $ \bnhadron = \Bd,\Bs ~{\rm or}~ \Lb$  and $h= K~{\rm or}~ \pi$) 
was reconstructed after the off-line confirmation of trigger requirements. 
In the offline analysis, an unbiased optimization procedure determined a
 tightened selection on track-pairs fit to a common decay vertex.
The selection cuts were chosen minimizing directly the expected uncertainty of the physics
observables to be measured (through several ``pseudo-experiments'').
Just two different sets of cuts were used in the analysis, 
respectively optimized to measure the \CP\ asymmetry \acpbdkpi\ (loose cuts) and to improve the 
sensitivity for discovery and limit setting~\cite{gp0308063} of the not yet observed \BsKpi\ mode (tight cuts).
For the \Lb\ measurements, the additional requirement $p_{T}(\bnhadron)>6~\pgev$ was applied to allow easy 
comparison with other \Lb\ measurements at the Tevatron, that are only available above this 
threshold \cite{lb_spectrum}.

\begin{figure}[htb]
\includegraphics[scale=0.35]{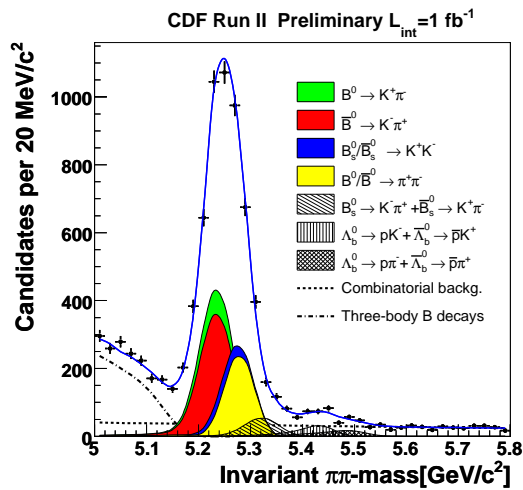}
\caption{Invariant mass distribution
of \bhh\ candidates passing tight cuts selection, 
using a pion mass assumption for both decay products.
Cumulative projections of the likelihood fit for each mode are
overlaid.}
\label{fig:projections}
\end{figure}

In addition to tightening the trigger cuts, in the offline analysis the discriminating power
of the \bnhadron\ isolation and of the information provided by the 3D reconstruction 
capability  of the CDF tracking were used,
allowing a great improvement in the signal purity.
 Isolation is defined as $I(\bnhadron)= \ptb/[\ptb + \sum_{i} \pt(i)]$, in which the sum runs over every other track 
(not from the $\bnhadron$ hadron) within
a cone of unit radius in the $\eta-\phi$ space around the \bnhadron\ hadron flight direction. 
By requiring $I(\bnhadron)> 0.5$,  the background was reduced by a factor
four while keeping almost 80\% of signal. The 3D silicon tracking allowed  multiple vertices  to be resolved
along the beam direction and the rejection of fake tracks, reducing the background
by a factor of two, with only a small efficiency loss on signal.
The resulting $\pi\pi$-mass distributions (see Figure.~\ref{fig:projections}) show a clean signal of \Bhh\ decays.
In spite of a good mass resolution ($\approx 22\,\massmev$), the various \bhh\ modes overlap into an unresolved
mass peak.

The resolution in invariant mass and in particle identification (\dedx) is not 
sufficient for separating the individual \Bhh\ decay modes on an event-by-event basis,
therefore a maximum likelihood fit was performed. This combines kinematic and particle identification information
to statistically determine both the contribution of each mode,
and the relative contributions to the \CP\ asymmetries.

Three separate fits were performed: one on the sample selected with loose cuts,  one
on the sample selected with tight cuts and one on the sample selected with tight cuts plus the requirement $\ptb>6~\pgev$. 
Significant signals are seen for \Bdpipi, \BdKpi, and \BsKK, previously observed by CDF~\cite{paper_bhh}.
Three new rare modes were observed for the first time \BsKpi, \Lbppi\ and \LbpK,
with a significance respectively of 
$8.2$, $6.0$ and $11.5$ Gaussian-equivalent standard deviation
estimated using a $p$-value distribution on pseudo-experiments. 
No evidence was obtained for \Bspipi or  \BdKK\ mode.
\begin{center}
\begin{table}[h]
\centering
\caption{\label{tab:summary} Branching fractions results. Absolute branching fractions are normalized to the the world--average values
${\mathcal B}(\mbox{\BdKpi}) = (19.4\pm 0.6) \times 10^{-6}$ and
$f_{s}/f_{d}= 0.276 \pm 0.034$ and $f_{\Lambda}/f_{d}= 0.230 \pm 0.052$~\cite{PDG08}.
The first quoted uncertainty is statistical, the second one is systematic.}
{
\begin{tabular}{|l|l|c|c|}
\hline
Mode          & Quantity & Measurement & \BR (10$^{-6}$)  \\
\hline
\Bdpipi        & \BdpipisuBdKpidef\  &  0.259 $\pm$ 0.017 $\pm$ 0.016   & 5.02 $\pm$ 0.33 $\pm$ 0.35 \\
\BsKK          & \BsKKsuBdKpidef\    &  0.347 $\pm$ 0.020 $\pm$ 0.021   & 24.4 $\pm$ 1.4 $\pm$ 3.5   \\
\hline
\BsKpi         & \BsKpisuBdKpidef\   &  0.071 $\pm$ 0.010 $\pm$ 0.007   & 5.0 $\pm$ 0.7 $\pm$ 0.8   \\
\LbpK          & \LbpKsuBdKpidef\    &  0.066 $\pm$ 0.009 $\pm$ 0.008   & 5.6 $\pm$ 0.8 $\pm$ 1.5  \\
\Lbppi         & \LbppisuBdKpidef\   &  0.042 $\pm$ 0.007 $\pm$ 0.006   & 3.5 $\pm$ 0.6 $\pm$ 0.9  \\
\hline
\Bspipi        & \BspipisuBdKpidef\  &  0.007 $\pm$ 0.004 $\pm$ 0.005   & 0.49 $\pm$ 0.28 $\pm$ 0.36 ($< 1.2$ @~90\%~CL) \\
 
\BdKK          & \BdKKsuBdKpidef\    &  0.020 $\pm$ 0.008 $\pm$ 0.006   & 0.39 $\pm$ 0.16 $\pm$ 0.12 ($< 0.7$ @~90\%~CL)  \\
\hline
\end{tabular}
}
\end{table}
\end{center}
\begin{center}
\begin{table}[h]
\centering
\caption{\label{tab:summary_ACP} CP asymmetries results. 
The first quoted uncertainty is statistical, the second one is systematic.}
{
\begin{tabular}{|l|c|}
\hline
Quantity & Measurement \\
\hline
\ACPddef\           & -0.086 $\pm$ 0.023 $\pm$ 0.009         \\
\ACPsdef\           &  0.39  $\pm$ 0.15  $\pm$ 0.08          \\
 
\ACPLbpKdef\        & -0.37  $\pm$ 0.17  $\pm$ 0.03          \\
\ACPLbppidef\       & -0.03  $\pm$ 0.17  $\pm$ 0.05          \\
\hline
\rateratiodef\      &  -3.00 $\pm$ 1.50 $\pm$ 0.22           \\
\hline
\end{tabular}
}
\end{table}
\end{center}

The relative branching fractions 
are listed in Table~\ref{tab:summary} while the CP asymmetries are listed in Table~\ref{tab:summary_ACP}, 
where $f_{d}$, $f_{s}$ and $f_{\Lambda}$ indicate
the production fractions respectively of \Bd, \Bs\ and \Lb\
from fragmentation of a $b$ quark in $\bar{p}p$ collisions.
An upper limit is also quoted for modes in which no significant signal is
observed. Absolute results are also listed in Table~\ref{tab:summary}, they are obtained 
by normalizing the data to the world--average of \BR(\BdKpi)~\cite{PDG08}. 

The branching fraction of the newly observed mode \BsKpi\ is in agreement with the latest
theoretical expectation \cita{zupan}, which is lower than  the previous predictions \cite{B-N,Yu-Li-Cai}.
This mode offers an unique opportunity to probe the direct CP violation in the \Bs\ mesons system.
For the first time, the direct \CP\ asymmetry in the \BsKpi\ decay mode was measured and its 
central value favors a large \CP\ violation (different from 0 at 2.3 Gaussian-equivalent standard deviation), 
although it is also compatible with zero. 
In Ref.~\cite{Lipkin:2005pb}  a robust test of the Standard Model or a probe of new physics is suggested by the
comparison of the direct \CP\ asymmetries in  \BsKpi\ and \BdKpi\ decays.
Using the external input for $f_{s}/f_{d}= 0.276 \pm 0.034$~\cite{PDG08} it is also possible to 
quote the following interesting quantity 
$\frac{\Gamma(\aBdKpi)-\Gamma(\BdKpi)}{\Gamma(\BsKpi)-\Gamma(\aBsKpi)}  
= 0.83 \pm 0.41 \pm 0.12$,  
which is in agreement with the Standard Model expectation of unity. Assuming this relationship true
(equal to unity), and using the external inputs for the \BR(\BsKpi), the world--average 
for the direct CP violating asymmetry in the \BdKpi\ decay mode~\cite{PDG08}, the \BR(\BdKpi)~\cite{PDG08}, 
it is possible to estimate the expected value for the direct CP violating asymmetry in the \BsKpi\ decay mode
($\approx 0.40$) which is in agreement with the CDF measurement presented here. 

The branching fraction of \BsKK\ decay mode 
is in agreement with the latest theoretical expectation~\cite{matiasBsKK,matiasBsKK2}
and with the previous CDF measurement \cite{paper_bhh}.

The results on the \Bd\ sector are in agreement with world--average values~\cite{PDG08}.
The direct CP violating asymmetry in the \BdKpi\ 
is competitive with the current \babelle\ measurements \cite{PDG08}.

The results on the \Lb\ sector are in agreement with Standard Model expectations.
The absolute branching fractions exclude $\mathcal{O}(10^{-4})$ values indicated for 
$R$-parity violating Minimal Supersymmetric extensions of the Standard Model~\cite{Mohanta2}.
The measurements of the direct CP violating asymmetries in the $b$-baryon decays, presented here,
are the first such measurements in this sector. The statistical uncertainty 
dominates the resolution and prevents a statement on the presence of asymmetry, whose 
measured value deviates from 0 at 2.1 Gaussian-equivalent standard deviation level in the 
\LbpK\ decay mode and is fully consistent with 0 in the \Lbppi\ decay mode. 

With full Run II samples ($5-6$~\lumifb\ by year 2010) CDF collaboration expects  a measurement of the direct CP violating asymmetry 
in the \BdKpi\ mode with a statistical plus systematic uncertainty at 1\% level; observation 
of the direct CP violating asymmetry in the \BsKpi\ mode (or alternatively the possible indication of non-SM sources 
of CP violation); more precise measurements of direct CP violating asymmetries in the \Lb\ charmless decays; and 
improved limits, or even observation, of annihilation modes \Bspipi\ and \BdKK.
In addition to the above, time-dependent measurements will be performed for \Bdpipi\ and \BsKK\ decay \cite{CKM06_punzi}. 
See \cite{beauty06_morello,my_thesis,Bhh_webpage,Lb_webpage} for more details.


\end{document}

%% file: macro.tex
\newcommand{\comment}[1]{}

\def \rightdownarrow
 {\kern.3em
 \rule[.5ex]{.15mm}{2ex}
 {\mbox{$\kern-0.1em{\longrightarrow}$}}
 }
\def\lessim{\mathrel {\vcenter {\baselineskip 0pt \kern 0pt
\hbox{$<$} \kern 0pt \hbox{$\sim$} }}}
\def\gessim{\mathrel {\vcenter {\baselineskip 0pt \kern 0pt
\hbox{$>$} \kern 0pt \hbox{$\sim$} }}}
\newcommand{\beq}{\begin{equation}}
\newcommand{\eeq}{\end{equation}}
\newcommand{\bear}{\begin{array}}
\newcommand{\ear}{\end{array}}
\newcommand{\bet}{\begin{tabular}}
\newcommand{\eet}{\end{tabular}}
\newcommand{\beqn}{\begin{eqnarray}}
\newcommand{\eeqn}{\end{eqnarray}}

\newcommand{\bfh}{\begin{figure}[h]}
\newcommand{\efh}{\end{figure}[h]}

\newcommand{\cita}[1]{\cite{#1}}

\newcommand{\tev}{\ensuremath{\mathrm{Te\kern -0.1em V}}}
\newcommand{\gev}{\ensuremath{\mathrm{Ge\kern -0.1em V}}}	
\newcommand{\mev}{\ensuremath{\mathrm{Me\kern -0.1em V}}}	
\newcommand{\kev}{\ensuremath{\mathrm{ke\kern -0.1em V}}}	
\newcommand{\massmev}{\mbox{\mev/$c^2$}}			
\newcommand{\pgev}{\mbox{\gev/$c$}}				

\newcommand{\CP}{CP}						
					
\newcommand{\pt}{\ensuremath{p_{\rm{T}}}}			
\newcommand{\ptb}{\ensuremath{\pt(\bnhadron)}}				

\newcommand{\bd}{\ensuremath{B^{0}}}				
\newcommand{\bs}{\ensuremath{B^{0}_s}}				
\newcommand{\bu}{\ensuremath{B^{+}}}				

\newcommand{\abd}{\ensuremath{\overline{B}^{0}}}		
\newcommand{\abs}{\ensuremath{\overline{B}^{0}_s}}		

\newcommand{\bhadrons}{\mbox{$b$-hadrons}}			

\newcommand{\bnhadron}{\mbox{$H^0_{b}$}}

\newcommand{\bhh}{\ensuremath{\bnhadron \to h^{+}h^{'-}}}

\newcommand{\bdkpi}{\ensuremath{\bd \to K^+ \pi^-}}

\newcommand{\etal}{{\em et al.}}

\newcommand{\Bd}{\ensuremath{B^{0}}}

\newcommand{\Bs}{\ensuremath{B_{s}^{0}}}
\newcommand{\Lb}{\ensuremath{\Lambda_{b}^{0}}}

\newcommand{\Bhh}{\ensuremath{\bnhadron \to h^{+}h^{'-}}}

\newcommand{\Bdpipi}{\ensuremath{\bd \to \pi^+ \pi^-}}

\newcommand{\BdKpi}{\ensuremath{\bd \to K^+ \pi^-}}
\newcommand{\aBdKpi}{\ensuremath{\abd \to K^- \pi^+}}
\newcommand{\BsKpi}{\ensuremath{\bs \to K^- \pi^+}}
\newcommand{\aBsKpi}{\ensuremath{\abs\to K^+ \pi^-}}
\newcommand{\BsKK}{\ensuremath{\bs \to  K^+ K^-}}

\newcommand{\Bspipi}{\ensuremath{\bs \to  \pi^+ \pi^-}}

\newcommand{\BdKK}{\ensuremath{\bd \to  K^+ K^-}}

\newcommand{\Lbppi}{\ensuremath{\Lambda_{b}^{0} \to p\pi^{-}}}
\newcommand{\aLbppi}{\ensuremath{\overline{\Lambda}_{b}^{0} \to \overline{p}\pi^{+}}}
\newcommand{\LbpK}{\ensuremath{\Lambda_{b}^{0} \to pK^{-}}}
\newcommand{\aLbpK}{\ensuremath{\overline{\Lambda}_{b}^{0} \to \overline{p}K^{+}}}

\newcommand{\BuKpi}{\ensuremath{\bu \to K^+ \pi^0}}

\newcommand{\Dkpi}{\ensuremath{D^{0} \to K^- \pi^+}}

\newcommand{\dedx}{\ensuremath{\rm{dE/dx}}}

\newcommand{\acpbdkpi}{\ensuremath{A_{\rm{\CP}}(\bdkpi)}}

\newcommand{\BR}{\ensuremath{\mathcal B}}
\newcommand{\cdf}{CDF Collaboration}

\newcommand{\babelle}{\mbox{$B$-Factories}}

\newcommand{\ACPddef}{\ensuremath{{\frac{\BR (\aBdKpi)-\BR 
(\BdKpi)}{\BR (\aBdKpi)+\BR (\BdKpi)}}}}
\newcommand{\ACPsdef}{\ensuremath{{\frac{\BR (\aBsKpi)-\BR 
(\BsKpi)}{\BR (\aBsKpi)+\BR (\BsKpi)}}}}

\newcommand{\ACPLbppidef}{\ensuremath{{\frac{\BR (\aLbppi)-\BR 
(\Lbppi)}{\BR (\aLbppi)+\BR (\Lbppi)}}}}
\newcommand{\ACPLbpKdef}{\ensuremath{{\frac{\BR (\aLbpK)-\BR 
(\LbpK)}{\BR (\aLbpK)+\BR (\LbpK)}}}}
\newcommand{\rateratiodef}{\ensuremath{\frac{\mathit{f_d}}{\mathit{f_s}}\times{\frac{ \Gamma(\aBdKpi)- 
\Gamma(\BdKpi)}{\Gamma(\aBsKpi)-\Gamma(\BsKpi)}}}}

\newcommand{\BdpipisuBdKpidef}{\ensuremath{\frac{\BR(\Bdpipi)}{\BR(\BdKpi)}}}
\newcommand{\BsKKsuBdKpidef}{\ensuremath{\frac{\mathit{f_s}}{\mathit{f_d}}\times\frac{\BR(\BsKK)}{\BR(\BdKpi)}}}

\newcommand{\BsKpisuBdKpidef}{\ensuremath{\frac{\mathit{f_s}}{\mathit{f_d}}\times\frac{\BR(\BsKpi)}{\BR(\BdKpi)}}}
\newcommand{\BspipisuBdKpidef}{\ensuremath{\frac{\mathit{f_s}}{\mathit{f_d}}\times\frac{\BR(\Bspipi)}{\BR(\BdKpi)}}}
\newcommand{\BdKKsuBdKpidef}{\ensuremath{\frac{\BR(\BdKK)}{\BR(\BdKpi)}}}

\newcommand{\LbppisuBdKpidef}{\ensuremath{\frac{\mathit{f_{\Lambda}}}{\mathit{f_d}}\times\frac{\BR(\Lbppi)}{\BR(\BdKpi)}}}
\newcommand{\LbpKsuBdKpidef}{\ensuremath{\frac{\mathit{f_{\Lambda}}}{\mathit{f_d}}\times\frac{\BR(\LbpK)}{\BR(\BdKpi)}}}

\newcommand{\Lumi}{\ensuremath{\mathcal{L}}}			
\newcommand{\lumifb}{\mbox{fb$^{-1}$}}				
